\documentclass[twocolumn,tighten,usenames,dvipsnames,twocolappendix]{aastex631}
\pdfoutput=1

\usepackage{amsmath}
\usepackage{amssymb}
\usepackage{amsthm}
\usepackage{MnSymbol}
\usepackage{nicefrac}
\usepackage{amsfonts}
\usepackage{dsfont}
\usepackage{hyperref}
\usepackage{natbib}
\usepackage{aas_macros} 
\usepackage{graphicx}
\usepackage{dcolumn}
\usepackage{bm}
\usepackage{color}
\usepackage{xcolor}
\newcommand{\bea}{\begin{eqnarray}}
\newcommand{\eea}{\end{eqnarray}}
\newcommand{\be}{\begin{equation}}
\newcommand{\ee}{\end{equation}}

\begin{document}

\title{
Horizonless spacetimes as seen by present and next-generation Event Horizon Telescope arrays
}
 
 \author{Astrid Eichhorn}
   \email{eichhorn@cp3.sdu.dk}
\affiliation{CP3-Origins, University of Southern Denmark, Campusvej 55, DK-5230 Odense M, Denmark}
 \author{Roman Gold}
   \email{gold@sdu.dk}
\affiliation{CP3-Origins, University of Southern Denmark, Campusvej 55, DK-5230 Odense M, Denmark}
\author{Aaron Held}
\email{aaron.held@uni-jena.de}
\affiliation{Theoretisch-Physikalisches Institut, Friedrich-Schiller-Universit\"at Jena, Max-Wien-Platz 1, 07743 Jena, Germany}
\affiliation{The Princeton Gravity Initiative, Jadwin Hall, Princeton University, Princeton, New Jersey 08544, U.S.}

\begin{abstract}
We study the capabilities of present and future radio very-long-baseline-interferometry arrays to distinguish black holes from horizonless spacetimes. We consider an example of a horizonless spacetime, obtained by overspinning a regular black hole. Its image is distinct from the image of a Kerr spacetime due to a second set of photon rings interior to the shadow. These photon rings cannot be directly resolved by present and even next-generation Event Horizon telescope arrays, but instead imprint themselves in horizon-scale images as excess central brightness relative to that of a black hole. We demonstrate that future arrays can detect such indirect imprints. 
\end{abstract}

\section{Introduction}
Is M87${}^*$ a black hole? In particular, does M87${}^*$ have a horizon? From a theoretical point of view, the distinction between a black hole and a horizonless object is a binary one -- the spacetime of a compact object is either in one or the other category. However, no observable directly encodes the presence of a horizon. Hence, one must resort to more indirect inferences. From an observational point of view, the distinction is thus  more gradual: imperfections of the observatory, such as finite resolution, limited uv-coverage and finite sensitivity, cause blurring, spurious features and limited dynamic range (roughly the difference of brightest to dimmest image features) in reconstructed images. 
However, with increased array performance, the dynamic range increases, allowing larger central brightness depressions to be detected. This enables  an improved distinction of horizonless objects and black holes (at least in the absence of astrophysical foreground emission).\\
Within the  uncertainties of the 2017 Event Horizon Telescope (EHT) observation, the image \citep{paper1, paper2, paper3, paper4, paper5, paper6} is compatible with a Kerr black hole and exhibits a central brightness depression. However, the dynamic range of the 2017 EHT observation is about 10, limiting the confidence with which it can be claimed that M87${}^*$ is indeed a black hole. Thus, it is paramount to investigate (i) whether horizonless objects could show a similar central brightness depression at 2017 EHT capabilities (see also \cite{EHT_SGRA_Paper6} (for Sgr A${}^*$) and \cite{Vincent:2020dij}) and (ii) quantify capabilities to constrain horizonless objects with current and future upgrades of the EHT array. 

To explore these questions, we consider a specific example of a horizonless spacetime. Its key characteristic, namely additional image structures interior to the shadow, are paradigmatic for a more general class of horizonless spacetimes \citep{Lamy:2018zvj,Rahaman:2021kge,Eichhorn:2022oma,Guerrero:2022msp,AEAH2022}.
We generate simulated high-resolution images of the spacetime illuminated by a simple accretion disk model.  These high-resolution images of horizonless objects differ from high-resolution images of black holes in several key characteristics. Chief among those are bright image features in the center. 
We use the ehtim toolkit~\citep{Chael:2018oym} to simulate how different very-long baseline interferometry (VLBI) arrays would reconstruct the image. This allows us to determine in how far any given VLBI array can distinguish horizonless objects from black holes.
In particular, we investigate how strongly the 2017 EHT observational campaign established the existence of a horizon and whether future next-generation EHT (ngEHT) campaigns will be able to distinguish horizonless objects from Kerr black holes.

\section{Horizonless spacetime metric}
We consider a spacetime given by the following line-element in ingoing Kerr coordinates $(u, r, \chi, \phi)$:
\begin{align}
\label{eq:metric}
ds^2 =&
-\frac{r^2-2M(r, \chi) r +a^2 \chi^2}{r^2+a^2 \chi^2}du^2 
+2\,du\, dr
\nonumber\\&
- 4\frac{M(r,\chi) a r}{r^2+a^2\chi^2}\left(1-\chi^2 \right) du\, d\phi
\nonumber
\end{align}
\begin{align}
&- 2a\left(1-\chi^2 \right)dr\, d\phi 
+ \frac{r^2+a^2\chi^2}{1-\chi^2}d\chi^2 
\nonumber\\&
+ \frac{1-\chi^2}{r^2+a^2\chi^2}\Big[
	\left(a^2+r^2\right)^2 
\nonumber\\&\quad
	- a^2\left(r^2-2M(r, \chi)r+a^2 \right) 
	\left(1-\chi^2 \right) 
\Big]d\phi^2,
\end{align}
where $a$ is the dimensionless spin parameter and the mass parameter $M$ is upgraded to a function of radius and azimuth,
\begin{align}
M(r, \chi) = 
\frac{M}{
	1+ \left[
		\ell_{\rm Planck}^4 \frac{48 M^2}{(r^2+a^2 \chi^2)^3}
		\right]^{1/2}.
}
\end{align}
Herein, $\ell_{\rm Planck}$ is a new-physics scale, which may, but need not be (even within quantum gravity theories) the Planck length, see, e.g.~\citet{Eichhorn:2022oma}. In the limit $\ell_{\rm Planck} \rightarrow 0$, the line element \eqref{eq:metric} corresponds to a Kerr black hole with asymptotic mass $M$ and angular momentum $J=a\,M$.
Deviations from the Kerr geometry are only sizable at (super-) Planckian curvature scales.
For
$a>a_{\rm crit} \approx M \left(1- 4\sqrt{3}\,\ell_{\rm Planck}^2/M^2 \right)+\mathcal{O}(\ell_{\rm Planck}^4)$, Eq.~\eqref{eq:metric} describes a horizonless spacetime.
This line element has several different possible interpretations: (i) as a regular, horizonless spacetime based on a locality-principle for new physics, as in \cite{Eichhorn:2021etc,Eichhorn:2021iwq}, (ii) as a spacetime inspired by asymptotically safe quantum gravity, pioneered in \cite{Bonanno:2000ep} and reviewed in \cite{AEAH2022}, which is a black hole for $a<a_{\rm crit}$ and a wormhole for $a>a_{\rm crit}$, (iii) as a Planck-scale regularized, overspun Kerr spacetime which corresponds to a wormhole geometry. Images of regular, overspun, horizonless spacetimes have been explored previously, e.g., in \cite{Lamy:2018zvj}, with a comparison to EHT data in \cite{Vincent:2020dij}; ngEHT capabilities have, to the best of our knowledge, not been explored in this context yet.

For the present purposes, the interpretation of Eq.~\eqref{eq:metric} is irrelevant.
We simply use it to sharpen the question, what the central brightness depression in the image of M87${}^*$ and the presence of a ring imply regarding the existence or non-existence of a horizon. Thus, we explore the line element as a paradigmatic example of horizonless spacetimes with (i) a photon sphere, (ii) negligible emission from within the central region, (iii) negligible absorption in the central region. The last two properties may be challenged from an astrophysical point of view: given that we are exploring a horizonless object with an accretion disk, one would expect (current and past) accretion which causes emission from within and absorption in the central region. Based on this expectation, arguments have been advanced that rule out horizonless spacetimes, see \cite{Broderick2015} (for M87${}^*$) and \cite{Broderick2009,Broderick2016} (for Sgr A${}^*$). Regarding these arguments, we discuss two caveats. First, the accreted matter is expected to reach the center of the spacetime, where the interplay of matter and spacetime within quantum gravity determines its ultimate fate. In the absence of an established theory of quantum gravity and matter, the ultimate fate of accreted matter and its absorption and emission profiles, are unknown. Second, if, e.g., Sgr A${}^*$ was a horizonless compact object, it must have an extremely low accretion rate, because otherwise thermal emission from the surface would be observable in the infrared \citep{Broderick:2005xa}. However, if the object is just a fraction more compact than the Schwarzschild radius, strong lensing leads to bouncing of photons on the hard surface, reducing detectable radiation significantly  \citep{Lu:2017vdx,Cardoso:2019rvt,Carballo-Rubio:2018jzw}. \\
For these reasons we perform a more {\it astrophysics-independent} investigation, i.e., one that does not rely on the assumptions -- as reasonable as they may be -- on the accretion history of the source.

\section{Disk model}
To test the observational distinguishability of this spacetime from a Kerr black hole, we need to account for the astrophysical environment, i.e., an accretion disk.
Instead of performing full GRMHD simulations\footnote{In the presence of new-physics effects in the spacetime metric, one may also expect new-physics effects in the matter equations. Accordingly, GRMHD simulations on top of a modified spacetime metric may not give access to all observational imprints of new physics.}, we work with a simple, non-dynamic emission model, which approximates the time-averaged structure arising from optically thin emission from an accretion disk, and integrate the radiative transfer equation with finite emissivity and vanishing absorptivity along null geodesics obtained by numerical ray tracing. We do not expect this simple model to account for a realistic astrophysical environment, but we expect that such emission provides a reasonable approximation to how spacetime features are illuminated.
The disk model is specified by a density function, cf.~\cite[Eq.~(2)]{Broderick:2021ohx},
\begin{align}
\label{eq:disk-model}
	n(r, \chi) = 
	n_0\,
	r^{-\alpha}e^{-\frac{\chi^2}{2\,h^2}}\,
	\begin{cases}
	0\;, & r<0 \\
	e^{-\frac{(r-r_\text{cut})^2}{\omega^2}}\;, & 0<r<r_\text{cut} \\
	1\;, & r>r_\text{cut}
	\end{cases}\;.
\end{align}
The rotation of the disk is specified by the fluid velocity $u_\mu = \bar{u}(-1,0,0,l)$ in Boyer-Lindquist coordinates with a polar angular momentum profile $l=R^{3/2}/(1+R)$ (where $R = r\sqrt{1-\chi^2}$), and the normalization $\bar{u}$ is chosen such that $u_\mu u^\mu\equiv-1$, cf.~\cite[Eq.~(6-8)]{Gold:2020iql}.
The exponential inner cutoff of the disk is set by the parameters $r_\text{cut}$ (radial location) and $\omega$ (sharpness): we choose $r_\text{cut}=4M$ and $\omega^2 = 1/12 \, M^2$. 
The large-distance falloff is set by the power-law exponent $\alpha$: we choose $\alpha = 3$.
The disk height along the black-hole spin axis is set by $h/r$. Our choice of $h=0.1$ crudely mimicks a time-averaged accretion flow in which the gas is confined to the near-equatorial regions as in magnetically arrested (or choked) disks \citep{McKinney2012,Narayan2003}.

We focus on the observability of interior image structures forming in horizonless spacetimes (without significant surface emission). In principle, varying the parameters of the disk model is also a way to obtain an estimate of astrophysical uncertainties, but we do not pursue this here given the simplicity of the disk model and the scope of this work.

\begin{figure*}[!t]
\centering
	\includegraphics[trim={3cm 18cm 3cm 18cm},clip,width=0.48\linewidth]{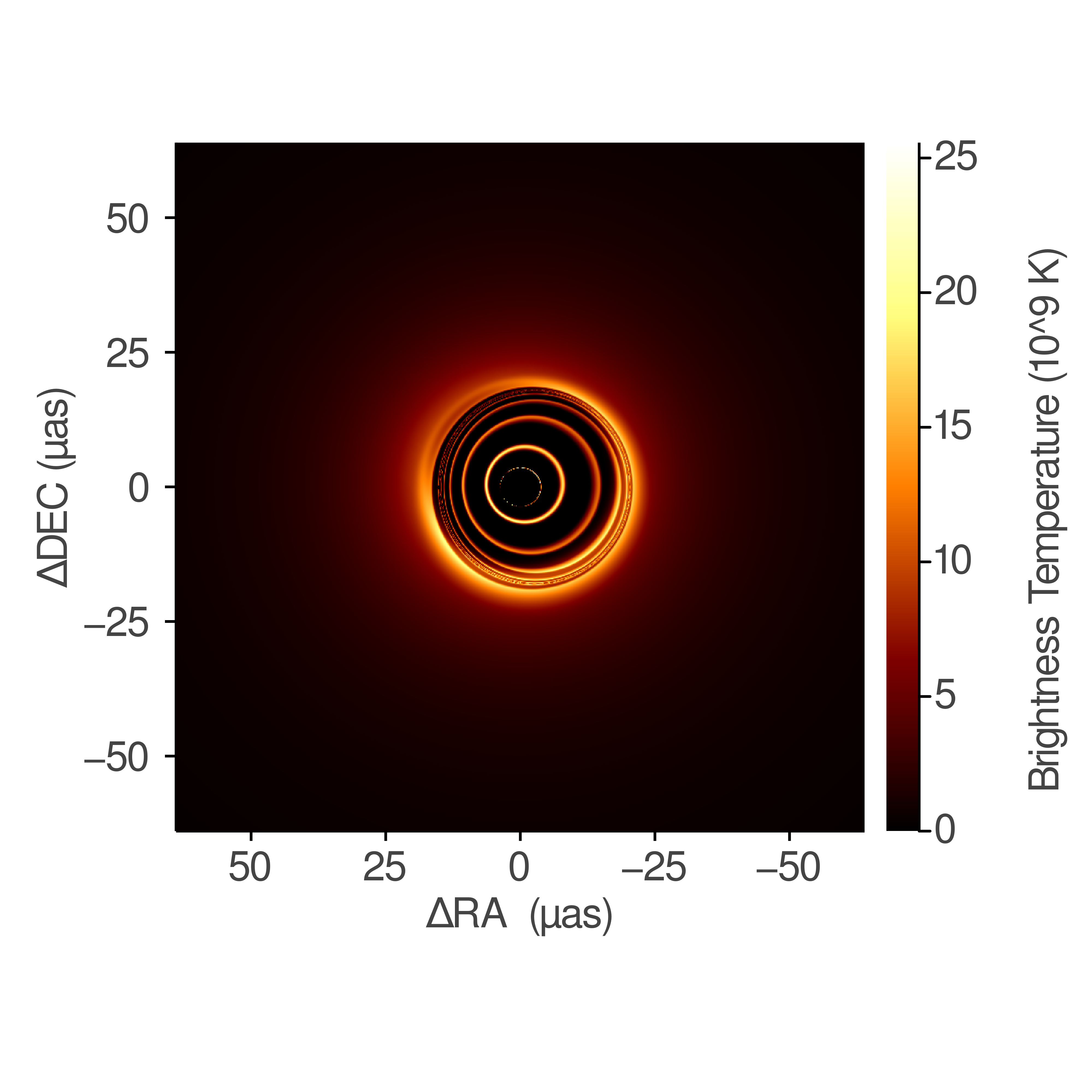}
	\hfill
	\includegraphics[trim={3cm 18cm 3cm 18cm},clip,width=0.48\linewidth]{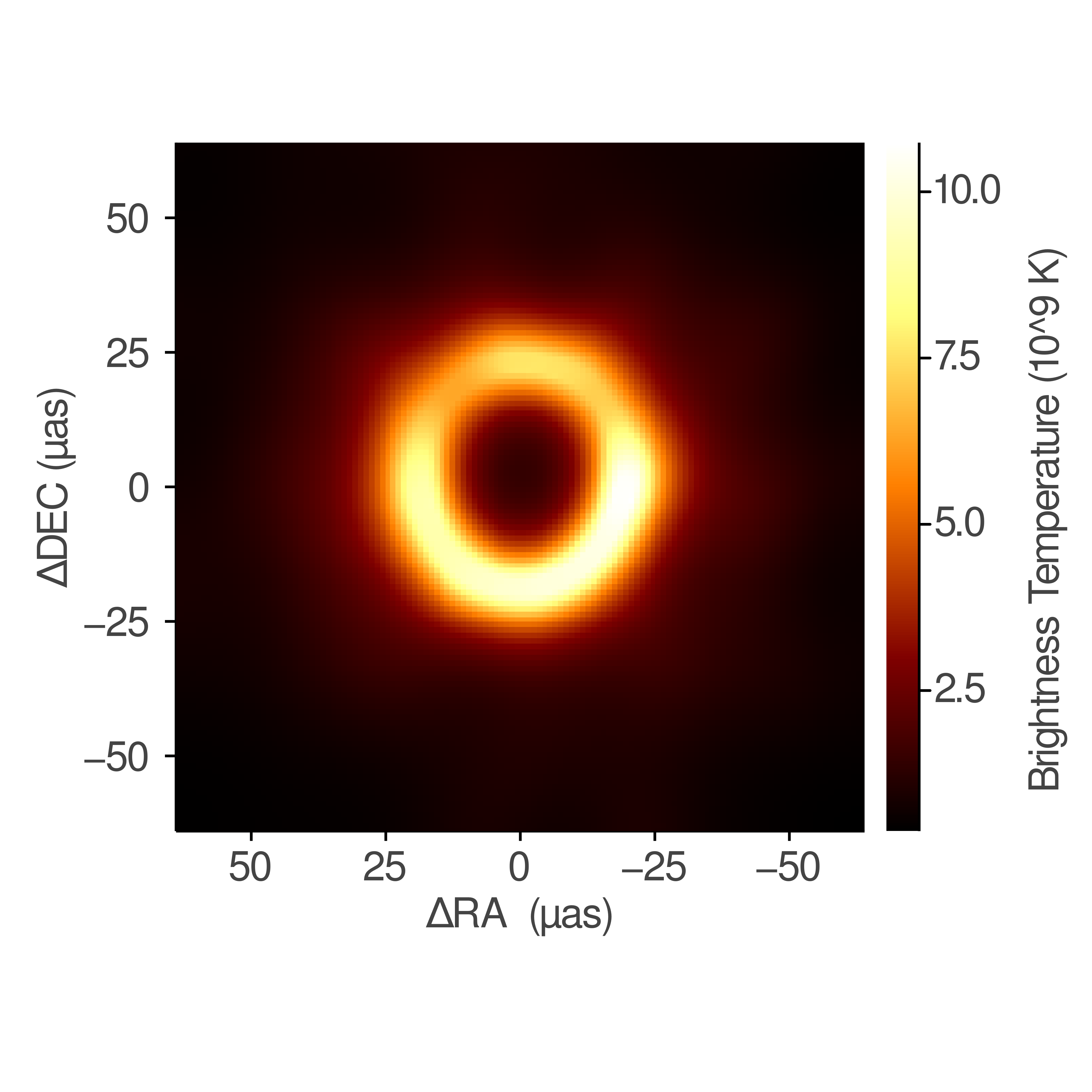}
\caption{\label{fig:horizonless}
We show the simulated image (left panel) and the reconstructed image (right panel) of a regular horizonless spacetime with $a=1.01$.
The simulated image is obtained by numerical ray tracing.
The reconstructed image is generated by the ehtim toolkit~\cite{Chael:2018oym} based on the EHT 2017 array at an observing frequency of 230 GHz.}
\end{figure*}

\section{Simulated image and inner photon rings}
We use numerical ray tracing (see~\cite{Held:2019xde,Eichhorn:2021iwq}
for details of our ray-tracing setup) to obtain a simulated image which constitutes the prediction of the model (spacetime + disk) and is distinct from an image of a Kerr black hole.

The distinctness lies in new lensing structures. These appear in the image region which features the shadow, i.e., a central brightness depression, in Kerr spacetime. The new lensing structures form a second set of photon rings, cf.~left panel in Fig.~\ref{fig:horizonless}, see also~\cite{Lamy:2018zvj,Rahaman:2021kge,Eichhorn:2022oma,Guerrero:2022msp,AEAH2022}. They are built up by null geodesics which start at large radii, approach the photon sphere from within and finally arrive in the image plane. In the presence of a horizon, those geodesics are intercepted by the horizon, such that a black-hole shadow is generated.
Just like for the outer photon rings, the inner photon rings are labelled by the number of times the geodesics pierce through the equatorial plane. The low-$n$ inner photon rings lie furthest inwards (just like the low-$n$ outer photon rings lie furthest outwards), and high-$n$ inner and outer photon rings approach each other more and more closely.

In the following, we specify the 
inclination $\theta_\text{obs}=17\pi/180$, as inferred from observations of the M87 jet~\citep{paper5}, and the BH spin parameter $a=1.01\,M$.
Any value of $a$ such that  $a_{\rm crit}<a\lesssim1.01 M$ will, for practical purposes, generate the same image, because differences require extremely high resolution to be visible. For values $a\gtrsim1.01\,M$, the interior image structures become coarser and thus easier to resolve, cf.~\cite{AEAH2022}. In this sense, we choose those parameters, $a$ and $M$, for which the image is least distinct from that of a Kerr spacetime.
The supercritical value of the spin parameter  does not result in a naked singularity, because the spacetime does not contain a curvature singularity. Whether or not regular black holes can be overspun by a physical process is a matter of debate, see, e.g., \cite{Li:2012ra,Li:2013sea,Jiang:2020mws,Yang:2022yvq} for studies, but cannot be answered solely within GR.

\begin{figure*}[!t]
\centering
	\begin{minipage}{0.7\linewidth}
	\includegraphics[width=\linewidth]{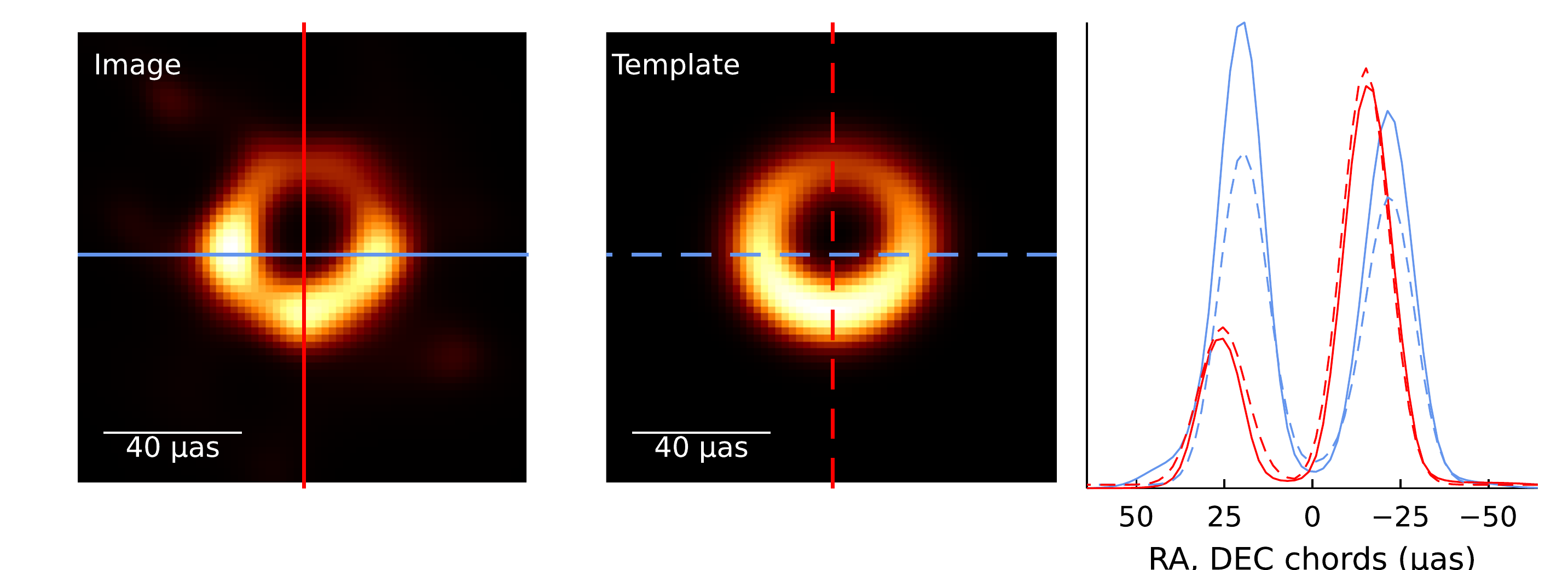}
	\end{minipage}
	\hfill
	\begin{minipage}{0.28\linewidth}
	\vspace*{6pt}
	\begin{flushleft}
	\begin{tabular}{r|c}
		{template parameter} & {value}
		\\\hline\hline
		$\sigma$ {(ring width [$\mu$as])} & 6.60
		\\
		$\tau$ {(asymmetry)} & 0.06
		\\
		$\xi_\tau$ {(asymm. orientation)} & 2.77
		\\
		$s$ {(slash)} & 0.47
		\\
		$\xi_s$ {(slash orientation)} & 4.89
		\\\hline
		$r_0$ {(ring radius [$\mu$as])} & 20.7
		\\
		$x_0$ {(offset RA [$\mu$as])} & -1.34
		\\
		$y_0$ {(offset DEC [$\mu$as])} & 5.29
		\\\hline\hline
		{ optimized divergence} & 0.017
	\end{tabular}
	\end{flushleft}
	\end{minipage}
	\\
	\begin{minipage}{0.7\linewidth}
	\includegraphics[width=\linewidth]{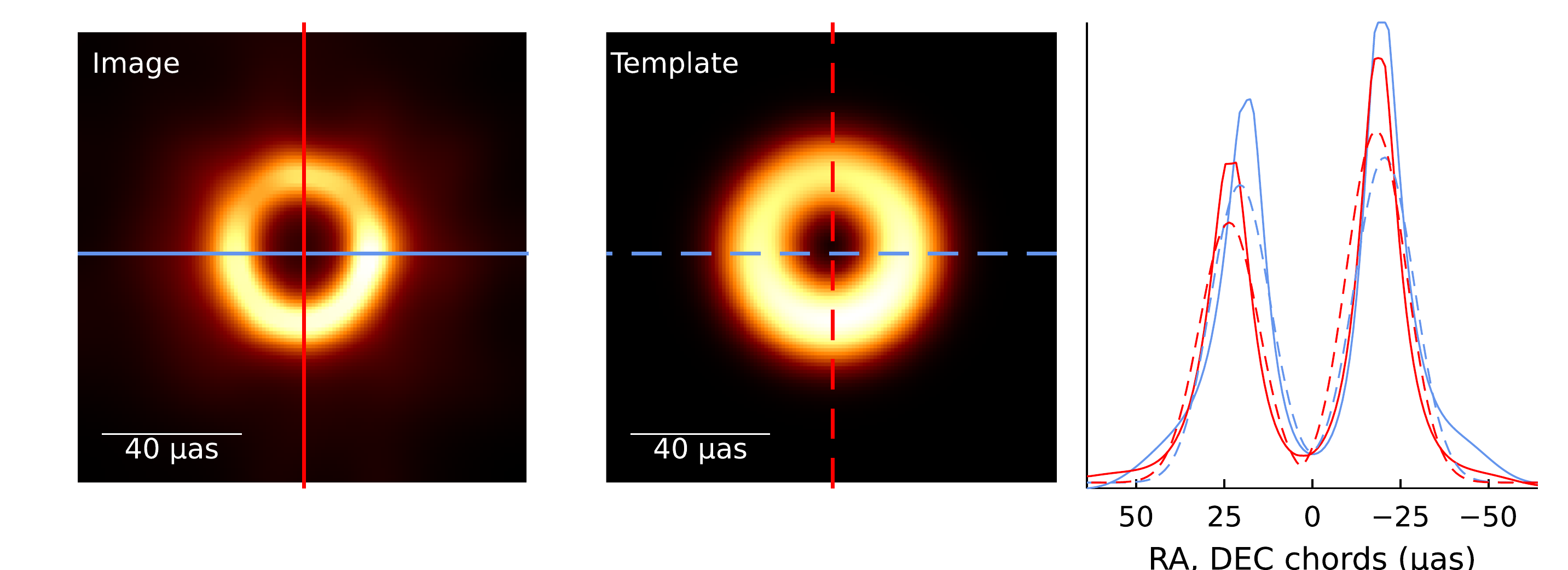}
	\end{minipage}
	\hfill
	\begin{minipage}{0.28\linewidth}
	\vspace*{6pt}
	\begin{flushleft}
	\begin{tabular}{r|c}
		{template parameter} & {value}
		\\\hline\hline
		$\sigma$ {(ring width [$\mu$as])} & 8.38
		\\
		$\tau$ {(asymmetry)} & 0.01
		\\
		$\xi_\tau$ {(asymm. orientation)} & 1.86
		\\
		$s$ {(slash)} & 0.157
		\\
		$\xi_s$ {(slash orientation)} & 4.42
		\\\hline
		$r_0$ {(ring radius [$\mu$as])} & 20.6
		\\
		$x_0$ {(offset RA [$\mu$as])} & 0.03
		\\
		$y_0$ {(offset DEC [$\mu$as])} & 2.73
		\\\hline\hline
		{ optimized divergence} & 0.006
	\end{tabular}
	\end{flushleft}
	\end{minipage}
\caption{\label{fig:M87fit}
We show the VIDA~\cite{Tiede:2020iif} fit of a GGR template (see main text) to (i) the M87${}^*$ image observed by the EHT on April 11 2017 (top panels) and to (ii) the reconstructed image of an overspun regular horizonless spacetime with spin $a=1.01\,M$ (bottom panel).
In each row, we show, from left to right, the ehtim-toolkit~\citep{Chael:2018oym}, reconstructed image (obtained with the 2017 EHT array configuration); the optimized template; the comparison of the image cross-sections of the reconstructed image (continuous) and the optimized GGR template (dashed); and a table of the optimized template parameters and divergence.
}
\end{figure*}

We investigate how the simulated image would be reconstructed by the EHT and how well it can be fit by relevant geometric templates.
Instead of fitting to EHT data and varying the free parameters of
  our model ($a$, $\alpha$, $h$, $r_{\rm cut}$ and $\omega$) to maximize
  fit quality, we work with an example image that is sufficiently similar
  to the M87${}^*$ image reconstructed by the EHT collaboration (EHTC). Apart from the
  similarity in terms of visual appearance in the image domain, our
  model also exhibits similar structures in the visibility amplitudes
  as a function of baseline, see Fig.~\ref{fig:uvdata}, and similar
  reconstructed features, see Fig.~\ref{fig:M87fit}. We consider it
  meaningful to demonstrate how (for a given disk model) parameter
  inference improves with array upgrades using the same method to
  measure features in the reconstructed images as done by the EHTC.
  In contrast, reproducing the astrophysics necessary to explain the 2017 EHT data at all image scales
  is not the point of this paper.  

\section{Reconstructed image}
We use the ehtim toolkit \citep{Chael:2018oym} to simulate the image reconstruction for a horizonless object located in M87 and observed by the 2017 EHT array \citep{paper2}. We neglect polarization and normalize the unpolarized intensity to the total image flux observed in the 2017 EHT April 11th observation.

We fix the overall angular size of the image in the observer's sky, which corresponds to setting the mass parameter $M$ to a specific value.
This uses the distance measurement of $16.8\pm0.8$~Mpc~\citep{Blakeslee:2009tc, 2010A&A...524A..71B, 2018ApJ...856..126C}.
To set the mass, we use the VIDA toolkit~\citep{Tiede:2020iif} to fit a general-Gau\ss ian-ring (GGR) template (see below) to the simulated image. We set units such that the diameter of the resulting ring agrees with the observed $42 \pm3\,\mu$as~\citep{paper1}.
This results in an asymptotic mass $M$ which is consistent with mass estimates from stellar dynamics \cite{Gebhardt:2011yw}.

The resulting reconstructed image is shown in the right panel of Fig.~\ref{fig:horizonless}. We report a fit value of $\chi^2 \approx 0.86$. This refers to the image reconstruction algorithm, which bases its $\chi^2$ on visibility amplitudes and closure phases as in \cite{paper4}.
The finite EHT resolution and dynamic range result in significant differences between the simulated  image (left panel) and the reconstructed image (right panel), as expected.
Specifically, a significant central brightness depression is present in the latter and the inner photon rings are unresolved. As a result, the reconstructed image shares many similarities with the observed image of M87${}^*$ in \cite{paper1,paper4}. A similar observation was made for the wormhole spacetime in  \cite{Vincent:2020dij}.

\begin{figure}[!t]
  \includegraphics[trim={0.3cm 0.3cm 1.5cm 1.3cm},clip,width=\linewidth]{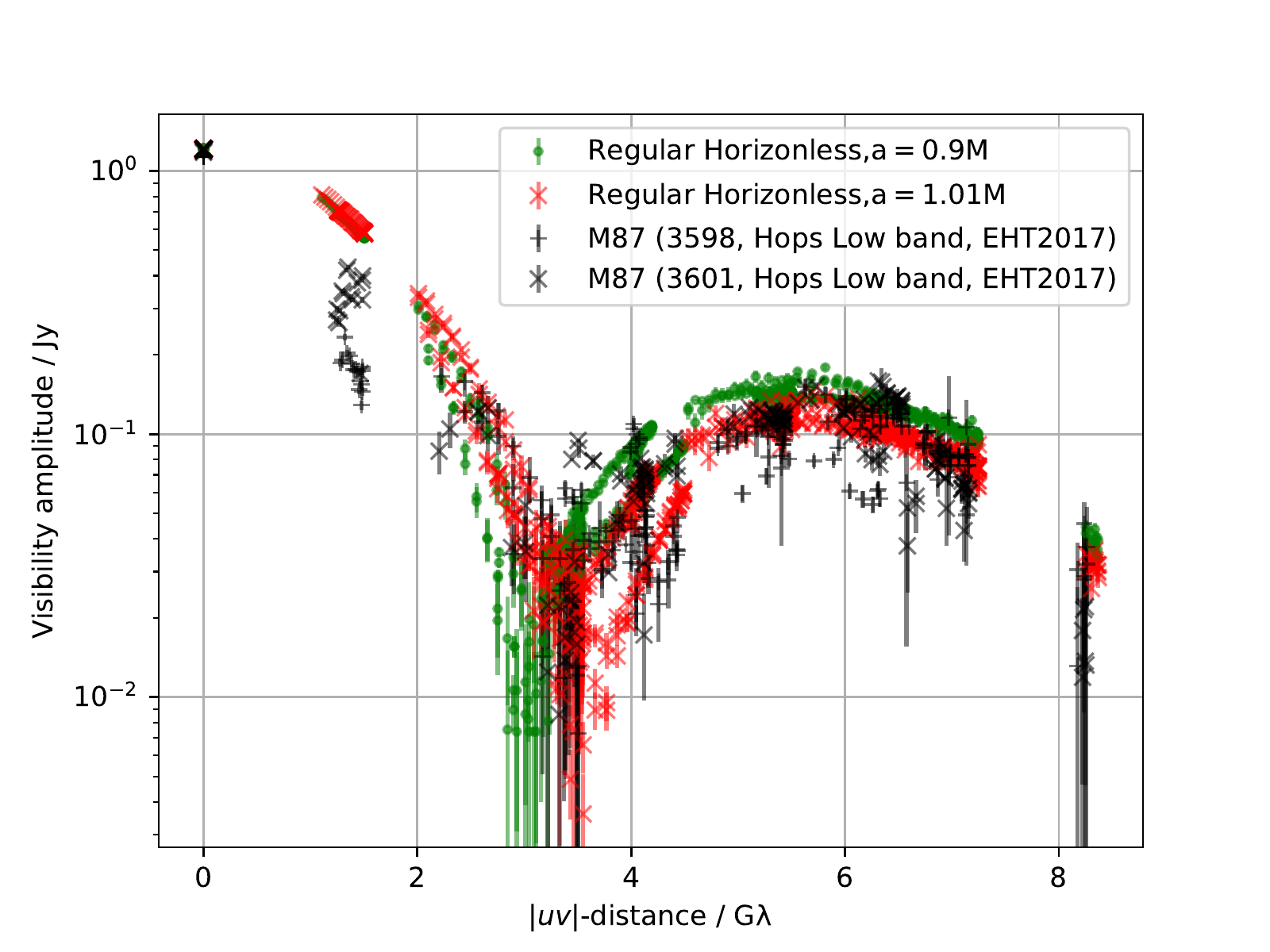}
\caption{\label{fig:uvdata} Comparison of visibility amplitudes as a
  function of $|\text{uv}|$-distance for the regular horizonless models with
  spin $a=0.9 M$ (green dots) and $a=1.01M$ (red crosses). Also shown are
  the EHT measurements from 3601 (Apr 11th) and 3598 (Apr 6th) from the 2017 EHT
  data, with thermal error bars~\citep{paper2,paper3}. At horizon scales ($|uv| >
  3\,\mathrm{G}\lambda$) the three cases are evidently quite similar,
  indicating that the model images produce similar features in the
  VLBI data as the actual measurements of M87. Larger scale structure
  is instead sensitive to astrophysics we do not model
  here.}
\end{figure}

As a next step, we go beyond the visual inspection and quantify the similarity in the following. 
We compare the visibility amplitude in the Fourier plane as a function of distance in the Fourier plane ($|\text{uv}|$-distance) to that of the 2017 EHT observing run of M87${}^*$ on April 6th and April 11th, see Fig.~\ref{fig:uvdata}. We observe that on both days, the data is in broad agreement with our simulated data above $|\text{uv}|$-distances of about 3 G$\lambda$. These large scales in the Fourier plane correspond to small scales in the image plane. The broad agreement confirms the visual inspection: at small scales, where additional inner photon rings exist in the simulated image, the reconstructed image closely resembles that of M87${}^*$. We further observe differences at smaller $|\text{uv}|$-distances, i.e., larger image scales, in particular for those distances where M87${}^*$ exhibits variability as becomes clear by comparing April 6 to April 11. This variability is of astrophysical origin, presumably due to dynamic jet emission structures seen at $\lambda=3mm$ \cite{Kim2018} and $\lambda=1mm$ \cite{Arras2020,Broderick2022inprep-RingsPaper}; thus we expect that either the present or an improved disk model would reach better agreement with the data at these small uv scales when disk parameters are varied. Our focus, however, is on large uv-scales, where properties of the spacetime dominate the image and which is therefore the relevant region for our comparison. To highlight that the difference at small $|\text{uv}|$-distances originates not in spacetime features, but in disk properties, we also show the case of a Kerr black holes with $a=0.9M$ with our disk model. It is in broad agreement with the EHT data and the horizonless spacetime at large $|\text{uv}|$-distances, but shows the same deviation from the EHT data at about 1.5 G$\lambda$ that our horizonless spacetime does.

\section{Template-based comparison between the EHT image of M87${}^*$ and the reconstructed image}
To quantitatively compare the reconstructed image 
of the regular horizonless spacetime with the image observed by the 2017 EHT observing run, we also perform a variational image-domain analysis using the VIDA toolkit~\citep{Tiede:2020iif}. VIDA compares a 2D image intensity $I(x,y)$ to a geometric  template $h(x,y)$ and thereby reduces the image information to a small set of template parameters. VIDA then optimizes the template parameters such that the Bhattacharyya divergence\footnote{VIDA uses both a Kullback-Leibler~\citep{Kullback:1951zyt} and a Bhattacharyya~\citep{Bhattacharyya1946} divergence. Both result in very similar optimized template parameters. Thus, we only report the results obtained by optimizing the Bhattacharyya divergence.}~\citep{Bhattacharyya1946}
\begin{align}
	D[I(x,y)||h(x,y)] = -\log\int\sqrt{I(x,y)h(x,y)}\,\text{d}x\text{d}y
\end{align}
between the image $I$ and the template $h$ is minimized. 

We use a general-Gau\ss ian-ring (GGR) template which suffices to extract the main image features of the observed EHT image, cf.~upper panels in Fig.~\ref{fig:M87fit}. The GGR template~\citep{Tiede:2020iif} is given by
\begin{align}
	\label{eq:GGR}
	h(x,y)\!=\!S(x,y;\,s,\xi_s)\exp\!\left[\frac{d_\theta(x,y;\,d_0,\tau,\xi_\tau,x_0,y_0)^2}{2\sigma^2}\right]\!.
\end{align}
Herein, $d_\theta(x,y;\,d_0,\tau,\xi_\tau,x_0,y_0)$ is the minimum distance of the respective image point at $(x,y)$ to an ellipse centered on $(x_0,y_0)$. 
The shape of the ellipse is parameterized by its mean diameter $d_0 = 2\sqrt{ab}$, asymmetry $\tau = 1-b/a$ (where $a$ and $b$ are the semi-major and semi-minor axis), and position angle $\xi_\tau$ of the semi-major axis (measured north of east). To account for a simple intensity asymmetry, the template also includes a slash function
\begin{align}
	S(x,y;\,s,\xi_s) = 1 + s\,\cos(\varphi - \xi_s)\;,
\end{align}
where $\varphi$ is the azimuthal image angle (again measured north of east). Overall the GGR template has 8 parameters: two coordinates $(x_0,y_0)$ that determine its center; the mean diameter $d_0$; the width $\sigma$; the degree and orientation $(\tau,\xi_\tau)$ of its asymmetry; and the degree and orientation $(s,\xi_s)$ of its slash.

We fit the template to the M87${}^*$ image from \cite{paper1}. The GGR template suffices to encode the relevant image features, as indicated by its small optimized divergence  
$D_\text{opt, M87}=0.017$.
Next, we fit the template to our reconstructed image and find a low optimized divergence 
$D_\text{opt, horizonless}=0.006$. Thus, the GGR template also provides a very good fit to our reconstructed image. However, unlike for the previous M87${}^*$ case, we do know that the underlying simulated image contains additional structures that are fundamentally mismatched to the choices made in the template. The good VIDA fit thus highlights that the 2017 EHT array is not capable of resolving the non-Kerr-like image features. Additionally, we find that most of the parameters of the fit, with the exception of the asymmetry orientation, are similar for the image of M87${}^*$ and our reconstructed image. This supports our conclusion based on comparing the visibility amplitudes in the Fourier domain. \\
Finally, the VIDA fit allows us to extract a quantitative value for the central brightness depression of the images using a consistent and well-defined procedure.

\section{Central brightness depression} 
We quantify whether one can tell apart the reconstructed and the observed image by the central brightness depression. We define the latter using the fitted GGR template: the ring region $\mathcal{R}$ is defined as the set of image points $(x,y)$ which are at most 1$\sigma$ away from the fitted ellipse, i.e.,
\begin{align}
	\label{eq:ringRegion}
	\mathcal{R}=\left\lbrace
		(x,y)\in\mathbb{R}^2\;:\;
		d_\theta(x,y;\,d_0,\tau,\xi_\tau,x_0,y_0)\leqslant\sigma
	\right\rbrace\;.
\end{align}
The shadow region $\mathcal{S}$ is defined as the interior complement of the ring. Following~\cite{paper6}, we  quantify the brightness ratio as
\begin{align}
	\label{eq:brightnessRatio}
	\hat{f}_{c} = \frac{\text{minimum flux in }\mathcal{S}}{\text{mean flux in }\mathcal{R}}\;,
\end{align}
evaluated on the reconstructed image (not the template fit). 

For the April 11 2017 M87${}^*$ image (see upper panel in Fig.~\ref{fig:M87fit} for the respective GGR template fit), we find a brightness ratio of $\hat{f}_{c}^\text{(M87${}^*$)} = 0.04 \ll 1$, i.e., a pronounced brightness depression. This is similar to the brightness ratios obtained in \cite{paper6} with other templates.

For the reconstructed image of the horizonless spacetime, we obtain a somewhat larger brightness ratio  of $\hat{f}_{c}^\text{(horizonless)} = 0.22$, i.e., a less pronounced brightness depression. 

These comparisons are independent of the precise definition of the brightness ratio: We vary the width of the ring region $\mathcal{R}$ by replacing $\sigma\rightarrow c\,\sigma$ in Eq.~\eqref{eq:ringRegion} with $c\in(0.1,2)$. For any value of $c\in(0.1,2)$, we find at least a factor of 5 between the two brightness ratios. Comparing to the EHT results is less straightforward, because (i) geometric modeling is fundamentally distinct from image reconstruction and (ii) measurements from reconstructions in \cite{paper4} did not use a template-based approach. With that being said, and in view of the combined modeling and observational uncertainties, the brightness ratios are broadly consistent with those reported in~\cite{paper4,paper6}. Note also that we did not actually fit to the data which would bring the flux ratios closer to the ones infered by the EHT.
\\

\begin{table}
	\begin{flushleft}
	\footnotesize
	\begin{tabular}{l|c|c}
		array configuration & $\hat{f}_c$ for $\frac{a}{M}=1.01$ & $\hat{f}_c$ for $\frac{a}{M}=0.9$
		\\\hline\hline
		EHT 2017 (230 GHz) & 0.216 & 0.070
		\\\hline
		EHT 2022 (230 GHz) & 0.156 & 0.036
		\\\hline
		ngEHT (230 GHz) & 0.159, 0.093* & 0.010
		\\\hline
		ngEHT (230 GHz multifreq) & 0.258 & 0.008 \\
		ngEHT (345 GHz multifreq) & 0.256 & 0.006
	\end{tabular}
	\end{flushleft}
	\caption{\label{tab:CBD}
	We tabulate the recovered brightness ratio $\hat{f}_c$ for a series of reconstructed observations with different array configurations.  
	See App.~\ref{app:VIDA230} for the value marked with *.}
\end{table}
\begin{figure*}[!t]
	\centering
	\includegraphics[width=\linewidth]{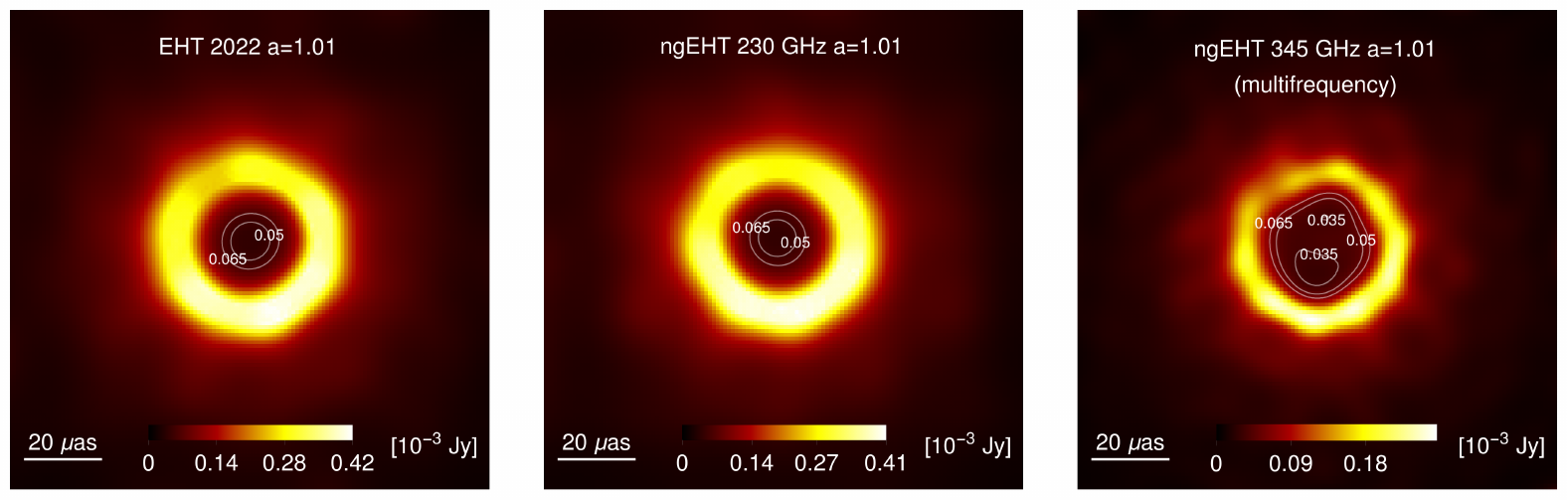}
\caption{\label{fig:ngEHT}
	We show reconstructed images of the regular horizonless spacetime with spin $a=1.01M$ as observed by different future array configurations and observing frequencies: The EHT 2022 array (10 telescopes
	operates at $230$ GHz (left panel). A potential ngEHT array configuration (20 telescopes) may observe at $230$ GHz (middle panel) but may also use a combined multifrequency observation at both $230$ GHz and $345$ GHz (right panel). We also show contour lines (thin white)
	to visualize small intensity differences in the interior.
}
\end{figure*}

\section{2022 EHT array and next-generation EHT}
Future VLBI arrays will benefit from (i) an increase in the number of telescope locations to increase the coverage in the Fourier plane, (ii) increased resolution when upgrading from $230$ GHz to $345$ GHz, and (iii) increased dynamic range, and can thus provide evidence against M87${}^*$ being a horizonless object. 

We consider the 2022 EHT array, to which several stations were added over those used in 2017. Further, we explore the capabilities of a next-generation EHT.
We add 10 new telescopes to the 2022 EHT-array and also explore the difference between an observation at $230$ GHz and an observation that combines data at $230$ GHz with data at $345$ GHz.
The location of the respective telescopes and their system equivalent flux density (SEFD) at both observing frequencies is (along with those of the 2017 and 2022 EHT  array configuration) given in Tab.~\ref{tab:telescopes} in App.~\ref{app:telescopes}, see also \cite{ngEHTexplorer.git}. We use a spectral index $\alpha=1.88$, as in \cite{EventHorizonTelescope:2021dvx}, resulting in a significantly smaller flux at 345~GHz than at 230~GHz.

We show the reconstructed images in Fig.~\ref{fig:ngEHT}, cf. Tab.~\ref{tab:fit-quality} in App.~\ref{app:telescopes} for the respective ehtim fit quality. The increased resolution has two important effects: The width of the ring shrinks for the ngEHT multifrequency case compared to the EHT observation. Moreover, the ngEHT (for the multifrequency case) is capable of detecting inner structure: contourlines at  low intensity
are concentric for the 2022 EHT array (left panel) and for the single-frequency ngEHT array (middle panel), but no longer concentric for the multifrequency ngEHT array (right panel). Instead, the center of the multifrequency ngEHT image is no longer the darkest spot. We conclude that the ngEHT is on the brink of being sensitive to the inner structure of the image of the investigated horizonless spacetime.

We again use a VIDA fit to the GGR template (see Tab.~\ref{tab:fit-quality} in App.~\ref{app:telescopes} for the respective optimized divergences) and calculate the brightness ratio, cf.~Tab.~\ref{tab:CBD}, where we compare to the brightness ratio of a Kerr black hole (with spin $a=0.9\,M$) with the same disk model. With increased dynamic range and resolution, the brightness ratio decreases for the Kerr black hole and is a factor of 10 smaller for the ngEHT multifrequency observation than for the EHT 2017 observation. In other words, for a Kerr black hole, the brightness depression is significantly more pronounced for the ngEHT than for the EHT reconstruction. 
In contrast, for a horizonless spacetime, there is no such trend; the central brightness depression does not become more pronounced for the ngEHT than for the EHT.
As a consequence, the difference in central brightness depression between Kerr and horizonless spacetime is at least one order of magnitude for the ngEHT case. 
 Under the assumption that both the model and the observational uncertainty on the brightness ratio decrease by an order of magnitude (such as the brightness ratio itself does between the 2017 EHT array and ngEHT array), the central brightness depression of a horizonless object and the Kerr spacetime differ with high significance.
\\
Therefore, if the ngEHT detects a significantly lower brightness ratio, i.e., less pronounced brightness depression than the 2017 EHT configuration, an increasingly strong case can be made against horizonless spacetimes. Conversely, if the ngEHT does not detect a  more pronounced brightness depression, no strong conclusion can be drawn, because the brightness depression can also be  reduced due to foreground emission (e.g., from a jet). However, if the  brightness depression is less pronounced due to foreground emission, one would expect to detect temporal variability over sufficiently long observational timescales, because a jet is not always radiating at the same intensity. In addition, it is evident from Fig.~\ref{fig:ngEHT} and the values for the flux ratios reported in Tab.~\ref{tab:CBD} that the central brightness depression is not capturing all of the significantly recovered features present in the reconstructed images for the future arrays. It must be expected that more complex templates can support a more sophisticated feature extraction that more effectively characterizes features present in the reconstructed images than we report here.
Therefore, we conclude that the prospects are promising for the ngEHT to better constrain horizonless spacetimes of the type we consider here. The use of two frequencies is particularly promising, because the multifrequency ngEHT reconstruction is on the brink of resolving the inner structure of the image of the horizonless spacetime, cf.~Fig.~\ref{fig:ngEHT}.

\section{Outlook: Radio-VLBI at super-resolution scales}
The extraction of image features is limited by the diffraction limit $\sim\lambda/D$ (with $\lambda$ the observing frequency and $D$ the baseline). With ground-based VLBI and a frequency of up to 345 GHz, the resolution is limited to $\sim$10 $\mu as$ and achieving higher resolution requires space-based VLBI, with its technological challenges \citep{Gurvits2022} and larger budget requirements as well as ecological costs. As an alternative, one can apply super-resolution techniques to ground-based VLBI data.

A promising method \citep{Broderick:2020wda} faithfully extracts the lowest order ($n=1$) photon ring  of a black hole from model images obtained via ray-tracing and synchrotron radiative transfer codes applied to GRMHD simulation data. The technique involves a mix of an image reconstruction plus a sharp ring model component that can model source structure below the diffraction limit, provided the signal-to-noise ratio is sufficiently high. When applied to the simulated data from emission models in the regular horizonless spacetimes investigated here, one expects this technique to also recover similarly sharp features allowing for stronger inferences of spacetime properties. We leave these investigations for future work.

\begin{acknowledgements} A.~E.~is supported by a research grant (29405) from VILLUM fonden. 
The work leading to this publication was supported by the PRIME programme of the
German Academic Exchange Service (DAAD) with funds from the German Federal Ministry of Education and Research (BMBF).
\end{acknowledgements}

\appendix

\begin{figure}[!t]
\centering
\includegraphics[width=0.7\linewidth]{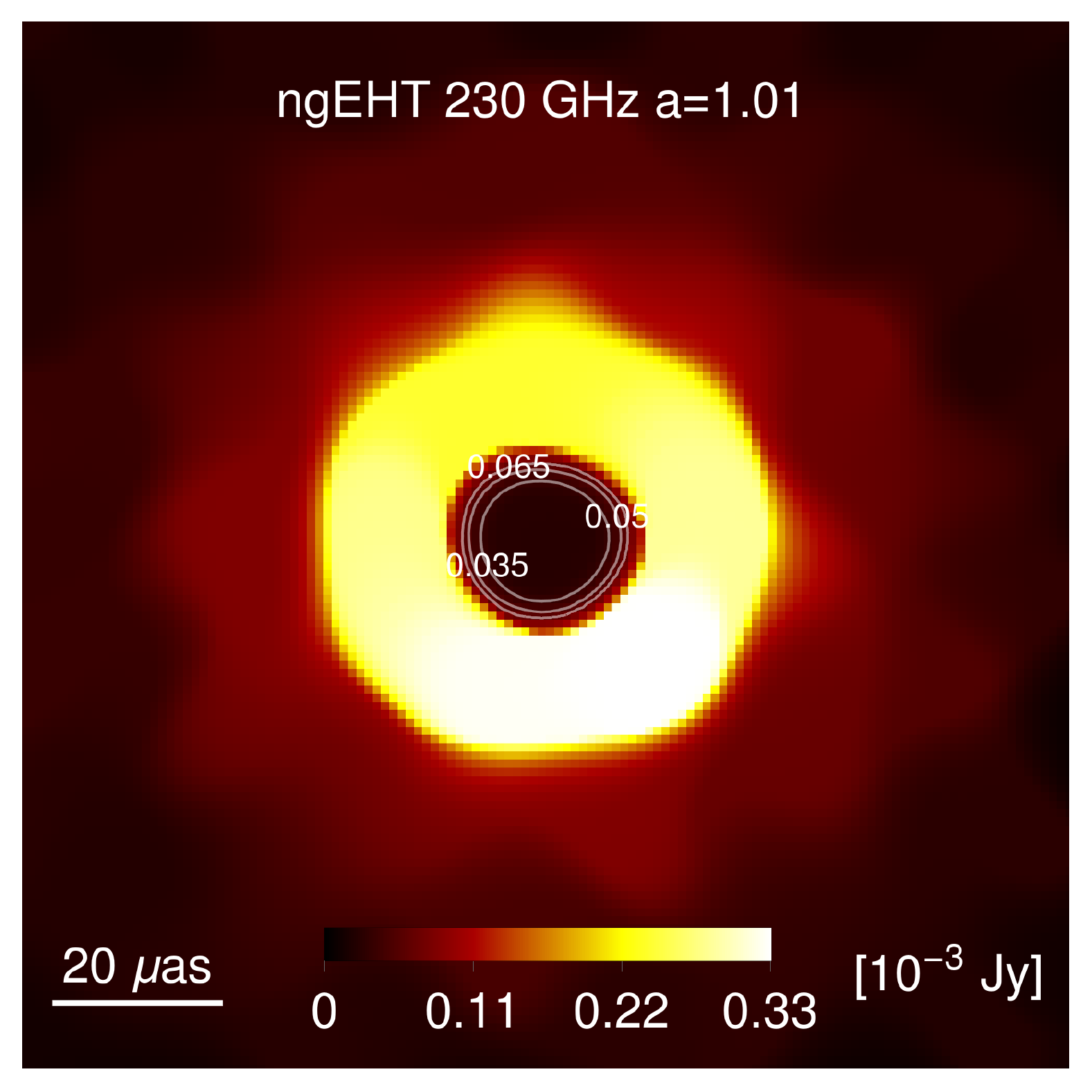}
\caption{\label{fig:secondimage} We show the second reconstruction of the simulated image by the ngEHT array at 230 GHz (see App.~\ref{app:VIDA230}).}
\end{figure}

\section{Telescope arrays}
\label{app:telescopes}

We detail the array configurations used when applying the ehtim toolkit \cite{Chael:2018oym} to reconstruct observations of the simulated image (cf.~left panel in Fig.~\ref{fig:horizonless}). The arrays are specified by a set of telescopes, their geocentric coordinates and their system equivalent flux density (SEFD) at 230 and 345 GHz, cf.~Tab.~\ref{tab:telescopes}.

We use three arrays: (i) the EHT 2017 array composed of 8 telescopes (cf. AA, AP, AZ, JC, LM, PV, SM, and SP in Tab.~\ref{tab:telescopes}, but note that the location of the South Pole Telescope (SP) prohibits any contributing baseline to an M87${}^*$ observation), (ii) the EHT 2022 array with the previous 8 and 3 additional telescopes (cf. GL, PB, and KP in Tab.~\ref{tab:telescopes}), and a potential ngEHT array with the previous 11 and 10 additional telescopes (cf. BA, BR, CI, GB, OV, SG, CT, GR, HA, and NZ in Tab.~\ref{tab:telescopes}). For the EHT 2017 and the EHT 2022 configuration, we reconstruct images at 230 GHz. For the ngEHT array, we reconstruct single-frequency images at 230 GHz as well as multifrequency images at 230 and 345 GHz.

\begin{table}
\begin{flushleft}
\footnotesize
	\begin{tabular}{l||r|r|r||r|r}
		 & $X_\text{geo}$ & $Y_\text{geo}$ & $Z_\text{geo}$ & $\text{\tiny SEFD}_{230}$ & $\text{\tiny SEFD}_{345}$  
		\\\hline\hline
		PV & 
		$5088968$ & 
		$-301682$ &  
		$3825016$ & 
		$330$& 
		$3850$
		\\
		AZ &
		$-1828796$ &
		$-5054407$ & 
		$3427865$ & 
		$2850$ &
		$17190$
		\\
		SM &
		$-5464523$ & 
		$-2493147$ & 
		$2150612$ & 
		$1230$ &
		$5730$
		\\
		LM &
		$-768714$ & 
		$-5988542$ & 
		$2063276$ & 
		$110$ &
		$2040$
		\\
		AA & 
		$2225061$ & 
		$-5440057$ & 
		$-2481681$ & 
		$40$ &
		$250$
		\\
		SP & 
		$0$ & 
		$0$ & 
		$-6359610$ &
		$7510$ &
		$25440$
		\\
		AP & 
		$2225040$ & 
		$-5441198$ & 
		$-2479303$ &
		$1790$ &
		$8880$
		\\
		JC & 
		$-5464585$ & 
		$-2493001$ & 
		$2150654$ & 
		$1190$ &
		$5780$
		\\\hline\hline
		GL & 
		$541647$ &
		$-1388536$ &
		$6180829$ &
		$4350$ &
		$14390$
		\\
		PB & 
		$4523999$ &
		$468045$ &
		$4460310$ &
		$300$ &
		$1410$
		\\
		KP & 
		$-1994314$ &
		$-5037909$ &
		$3357619$ &
		$7430$ &
		$44970$
		\\\hline\hline
		BA & 
		$-2352576$ & 
		$-4940331$ & 
		$3271508$ & 
		$16930$& 
		$58500$
		\\
		BR & 
		$-2363000$ & 
		$-4445000$ & 
		$3907000$ & 
		$15770$ & 
		$52160$
		\\
		CI & 
		$5311000$ & 
		$-1725000$ & 
		$3075000$ & 
		$19410$ & 
		$76110$
		\\
		GB & 
		$5627890$ & 
		$ 1637767$ & 
		$-2512493$ & 
		$14270$ & 
		$264200$
		\\
		OV & 
		$-2409598$ & 
		$-4478348$ & 
		$3838607$ & 
		$15100$ & 
		$118890$
		\\
		SG & 
		$1832000$ & 
		$-5034000$ & 
		$-3455000$ & 
		$17570$ & 
		$63760$
		\\
		CT & 
		$1569000$ & 
		$-4559000$ & 
		$-4163000$ & 
		$29580$ & 
		$167890$
		\\
		GR & 
		$1538000$ & 
		$-2462000$ & 
		$-5659000$ & 
		$71080$ & 
		$736630$
		\\
		HA & 
		$1521000$ & 
		$-4417000$ & 
		$4327000$ & 
		$2740$ & 
		$66530$
		\\
		NZ & 
		$-4540000$ & 
		$719000$ & 
		$-4409000$ & 
		$32040$ & 
		$191080$
	\end{tabular}
	\caption{\label{tab:telescopes}
	We tabulate the geocentric coordinates $(X_\text{geo},Y_\text{geo},Z_\text{geo})$ (in units of $m$) and the system equivalent flux density (SEFD) at $230$ and $345$ GHz (in units of Jy). The first 8 lines denote the EHT 2017 telescopes. The next 3 lines denote the telescopes added to the EHT 2022 array. The final 10 lines denote a choice of telescopes tentatively added to a future ngEHT array, cf.~\cite{ngEHTexplorer.git}.
	}
\end{flushleft}
\end{table}

In order to compare the central brightness depression of black-hole ($a=0.9\,M$) and horizonless ($a=1.01\,M$) spacetimes for each of the above four array configurations, we repeat the analysis detailed in the main text for the EHT 2017 array and the horizonless $a=1.01\,M$ spacetime. In Tab.~\ref{tab:fit-quality}, we provide the respective $\chi^2$ values of the ehtim reconstruction and the optimized divergence value for the fitted VIDA GGR template.

\begin{table}
	\begin{flushleft}
	\footnotesize
	\begin{tabular}{l|l|c|c}
		$a$ & array configuration & $\chi^2_\text{ehtim}$ & $D_\text{opt}$
		\\\hline\hline
		$1.01$ & EHT 2017 (230 GHz) & 0.92 & 0.007
		\\\cline{2-4}
		 & EHT 2022 (230 GHz) & 0.94 & 0.009
		\\\cline{2-4}
		 & ngEHT (230 GHz) & 0.98, 0.97* & 0.009, 0.012*
		\\\cline{2-4}
		 & ngEHT (230 GHz multifreq) & 0.99 & 0.012
		\\
		& ngEHT (345 GHz multifreq) & 0.98 & 0.013
		\\\hline\hline
		$0.9$ & EHT 2017 (230 GHz) & 0.96 & 0.011
		\\\cline{2-4}
		 & EHT 2022 (230 GHz) & 0.94 & 0.018
		\\\cline{2-4}
		 & ngEHT (230 GHz) & 1.01 & 0.027
		\\\cline{2-4}
		 & ngEHT (230 GHz multifreq) & 0.99 & 0.028
		\\
		 & ngEHT (345 GHz multifreq) & 0.99 & 0.032
		\\
	\end{tabular}
	\end{flushleft}
	\caption{\label{tab:fit-quality}
	We tabulate the $\chi^2$ value for the ehtim image reconstruction ($\chi^2_\text{ehtim}$) as well as the optimized divergence value for the VIDA GGR template fit ($D_\text{opt}$) for all the discussed reconstructed observations and values of the spin parameter $a$ (in units of $M$).
	See App.~\ref{app:VIDA230} for the values marked with *.
	}
\end{table}

\section{Image reconstruction and VIDA fit at 230 GHz ngEHT}\label{app:VIDA230}
In addition to the central panel in Fig.~\ref{fig:ngEHT}, which is reconstructed with $\chi^2=0.98$, and for which we obtain a brightness ratio of $\hat{f}_c=0.156$, a second image is reconstructed with a very similar fit quality of $\chi^2=0.97$ but rather distinct image morphology, cf.~Fig.~\ref{fig:secondimage}. Evidently, for this case, the likelihood surface is multimodal and more difficult to traverse. The latter fit achieves a slightly better fit quality by driving up the regularization term causing excessive smoothing to the reconstructed image resulting in a ring with much larger width than in the simulated image or any of the other reconstructions. Here, we have the advantage of knowing the input simulated image and can make this assertion. For an actual observation, this situation would be difficult to assess. In the interest of a conservative result, and for transparency, we carry out the analysis for both fits including the excessively smoother one. Using the optimized VIDA GGR template (with $D_\text{opt}=0.012$, see main text for details) to quantify the central brightness depression, we obtain a brightness ratio of $\hat{f}_c=0.093$. Comparing the two values of $\hat{f}_c$, obtained from the same simulated image, but the two different reconstructions, we can estimate a lower bound on the error in $\hat{f}_c$ for this given array configuration, i.e., $\Delta \hat{f}_c \approx 0.06$.

\bibliography{References}
\bibliographystyle{aasjournal}

\end{document}